\documentclass[%
 amsmath,amssymb,
 aps,
 longbibliography,
]{revtex4-1}

\usepackage{graphicx}% Include figure file
\usepackage{dcolumn}% Align table columns on decimal point
\usepackage{bm}% bold math
\pdfoutput=1
\renewcommand{\toprule}{\hline}
\newcommand{\midrule}{\hline}
\newcommand{\bottomrule}{\hline}
\usepackage{lineno}
\usepackage{datatool-base}

\begin{document}

\preprint{ILD}
\rightline{DESY 19-215}

\title{The ILD Detector at the ILC}% Force line breaks with \\

%\affiliation{Deutsches Elektronen Synchroton, DESY, Germany}

\collaboration{The ILD Collaboration\footnote{corresponding author: Ties Behnke, DESY}}%\noaffiliation
%\author{Contact: \DTLinitials{Ties} Behnke, Deutsches Elektronen Synchroton, DESY}

\begin{abstract}
The International Large Detector, ILD, is a detector concept which has been developed for the electron-positron collider ILC. The detector has been optimised for precision physics in a range of energies between 90~GeV and 1~TeV. ILD features a high precision, large volume combined silicon and gaseous tracking system, together with a high granularity calorimeter all inside a 3.5\,T solenoidal magnetic field. The paradigm of particle flow has been the  guiding principle of the design of ILD. In this document the required performance of the detector, the proposed implementation and the readiness of the different technologies needed for the implementation are discussed. This is done in the framework of the ILC collider proposal, now under consideration in Japan, and includes site specific aspects needed to build and operate the detector at the proposed ILC site in Japan.

%\begin{description}
%\end{description}
\end{abstract}

%\author{	\DTLinitials{Halina}	Abramowicz	}
%\author{	\DTLinitials{Ishikawa}	Akimasa	},
%\author{	\DTLinitials{Miyamoto}	Akiya	},
%\author{	\DTLinitials{Mohammad Sohail}	Amjad},
%\author{	\DTLinitials{Dieguez}	Angel	},
%\author{	\DTLinitials{Justin}	Anguiano	}

\author{\DTLinitials{Halina} Abramowicz$^{50}$}
\author{\DTLinitials{Tatjana} Agatonovic Jovin$^{61}$}
\author{\DTLinitials{Oscar} Alonso$^{52}$}
\author{\DTLinitials{Mohammad Sohail} Amjad$^{53}$}
\author{\DTLinitials{Fenfen} An$^{20}$}
\author{\DTLinitials{Ladislav} Andricek$^{15}$}
\author{\DTLinitials{Marc} Anduze$^{34}$}
\author{\DTLinitials{Justin} Anguiano$^{28}$}
\author{\DTLinitials{Evgeny} Antonov$^{36}$}
\author{\DTLinitials{Yumi} Aoki$^{29}$}

% ...
\affiliation{$^{15}$Halbeiterlabor der Max-Planck-Gesellschaft, M\"unchen, Germany}
% ...
\affiliation{$^{20}$Institute of High Energy Physics, Chinese Academy of Science, Beijing, China}
% ...
\affiliation{$^{28}$University of Kansas, Department of Physics and Astronomy, Lawrence, KS, USA}
\affiliation{$^{29}$High Energy Accelerator Research Organisation, KEK, Tsukuba, Ibaraki, Japan}
% ...
\affiliation{$^{34}Laboratoire Leprince-Ringuet (LLR), $CNRS/IN2P3, Ecole Polytechnique, Institut Polytechnique de Paris, Palaiseau, France}
% ...
\affiliation{$^{36}$P.N. Lebedev Physical Institute of the Russian Academy of Sciences  (LPI), Moscow, Russia}
% ...
\affiliation{$^{50}$Tel Aviv University, Raymond \& Beverly Sackler School of Physics \& Astronomy, Tel Aviv, Israel}
% ...
\affiliation{$^{52}$University of Barcelona, Barcelona, Spain}
\affiliation{$^{53}$University College London, London, UK}
% ...
\affiliation{$^{61}$VINCA Institute of Nuclear Sciences, University of Belgrade, Belgrade, Serbia
 \vspace{3mm}}

% ...

%\author{ILD author list}
% \homepage{http://www.ilcild.org}
%\affiliation{
%\affiliation{
%\affiliation{%

%\collaboration{ILD Collaboration}%\noaffiliation

\date{\today}% It is always \today, today,


%\begin{center}

%\includegraphics[width=0.8\hsize]{figures/ILD.pdf}
%\end{center}

\maketitle

%\tableofcontents

%\newpage

\section{\label{sec:level1}Introduction}
The International Large Detector, ILD, is a detector proposal for the International Linear Collider, ILC. In this paper, the considerations which have guided the ILD concept group in the design of the detector are summarized. The main technological challenges for the realisation of the concept are described, and possible technological solutions are sketched. The ILD concept is supported by a broad and international community of scientists. For the development of the concept particular emphasis has been put on the realism of the simulation model used, both for optimisation studies and for physics benchmarking.

The ILD detector concept group was formed in 2007, as a merger of two earlier detector concepts, GLD~\cite{ild:bib:ref-gld} and LDC~\cite{ild:bib:ref-ldc}. GLD was a concept for the Asian Linear Collider, LDC for the TESLA linear collider proposal. Both detector concepts were similar in that they relied on a combination of silicon and gaseous tracking, combined with precision calorimetry, though in detail the solutions were rather different. Following the agreement by the international community to continue with only one linear collider concept, the ILC, the two concepts joined forces. During 2007 and 2008, an intense effort took place to define the new detector based on the work done in the two previous concept groups. 

At about the same time, particle flow as a novel idea to reconstruct complex events at a collider had become more generally accepted - although a convincing experimental study that the required resolution could be reached was still missing at that time. ILD decided nevertheless to adopt particle flow as the central guiding principle for its detector concept, and developed the ILD design around this paradigm. For a review on particle flow, see e.g. \cite{ild:bib:PandoraPFA}.

The ILD concept underwent a number of international reviews, and was validated as one of two ILC detector concept groups. After the delivery of the ILD detector baseline document in 2013~\cite{Behnke:2013lya}, ILD re-organized its structure, to respond to the increasing possibility that ILC as a project would be realized in Japan. An optimization process was started to react to new technical developments, as well as to the quest for reducing the cost of the overall project. The detailed design considerations, performance studies and cost updates will be available in the ILD design report, which is under preparation. 

\section{The ILD detector design: requirements}
The science which will be done at the ILC has been summarized in a separate document~\cite{ILCESU1}. It is strongly dominated by the quest for ultimate precision in measurements of the properties of key particles like the Higgs boson, the weak gauge bosons, and, once the center-of-mass energy is beyond its production threshold, the top quark (see for example~\cite{Fujii:2017vwa} for a recent summary). 

%A more detailed description of the ILD detector and its design philosophy is available in \cite{Behnke:2013lya}.

The anticipated precision physics program drives the requirements for the detector. Many final states which will be analysed are hadronic final states, with many jets. Thus a precise reconstruction of jets is essential, which translates into an excellent jet energy resolution. Several studies that investigated the reconstruction of $W$ and $Z$ bosons suggest that a jet energy resolution of about 3\% is needed to fully exploit the power of the collider. Such a resolution requires an improvement of performance; it is almost two times more precise than the ATLAS and CMS detectors at the LHC. The concept of particle flow is currently believed to be the only practical approach which can reach this level of precision. Particle flow requires the reconstruction of charged and neutral particles with excellent efficiency over a large solid angle, although not requiring an exceptional  resolution for individual particles. Thus a tracker with outstanding efficiency is needed, combined with a calorimeter capable of reconstructing neutral particles with high efficiency. For ILD the choice has been made to combine a large volume gaseous tracking system - which promises excellent efficiency combined with low material - and a highly granular calorimeter both in the electromagnetic and the hadronic sections. To ease linking between the tracker and the calorimeter, the calorimeter should be inside the coil. The very high granularity proposed requires significant power consumption at the detector front-end. For ILC this is manageable since the bunch-train structure of the ILC allows the electronics to be power pulsed, with a duty cycle (ratio total time to on time) between 10 and 100. This reduces the average power consumption to a level where in most cases no or only very little active cooling is needed at the actual detector. 

A number of highly relevant physics processes require the precise reconstruction of exclusive final states containing heavy flavor quarks. This translates into the need for very precise reconstruction of the decay vertices of long lived particles, and thus implies a high resolution vertexing system close to the interaction region. 
The excellent performance of the tracking system depends critically on the amount of material in the inner part of the ILD detector. The total material budget in front of the calorimeter should be below 10\% of a radiation length, for the barrel part of the detector acceptance.

The show-case reaction of the recoil $HZ$ analysis requires high precision tracking, to be able to reconstruct the di-muon decay of the $Z$ boson, against which the Higgs recoils, with a precision not limited by detector resolution effects. This adds excellent momentum reconstruction precision to the list of requirements. 

The design drivers of the ILD detector can be summarized by the following requirements: 
\begin{itemize}
    \item {\bf Impact parameter resolution:}  An impact parameter resolution of $ 5~\mu \mathrm{m} \oplus 10~\mu \mathrm{m} / [ p~({\mathrm{GeV}/c})\sin^{3/2}\theta$] has been defined as a goal, where $\theta$ is the angle between the particle and the beamline. 
    \item {\bf Momentum resolution:} An inverse momentum resolution of $\Delta (1 / p) = 2 \times 10^{-5}~\mathrm{(GeV/c}^{-1}/$) asymptotically at high momenta should be reached with the combined silicon - TPC tracker. Maintaining excellent tracking efficiency and very good momentum resolution at lower momenta will be achieved by an aggressive design to minimise the detector's material budget.
    \item {\bf Jet energy resolution:} Using the paradigm of particle flow a jet energy resolution $\Delta E/ E = 3\%$ for light flavour jets should be reached. The resolution is defined in reference to light-quark jets, as the R.M.S. of the inner $90\%$ of the energy distribution. 
    \item {\bf Readout:} The detector readout will not use a hardware trigger, ensuring full efficiency for all possible event topologies.
    \item {\bf Powering} To allow a continuous readout, and, at the same time, minimize the amount of dead material in the detector, the power of major systems will be cycled between bunch trains. 
\end{itemize}
%A quadrant view of the ILD detector is shown in figure~\ref{fig:ILD} (left).

\begin{figure}[tb]
 \begin{center}
 \begin{tabular}{lr}
 \includegraphics[width=0.48\hsize,clip]{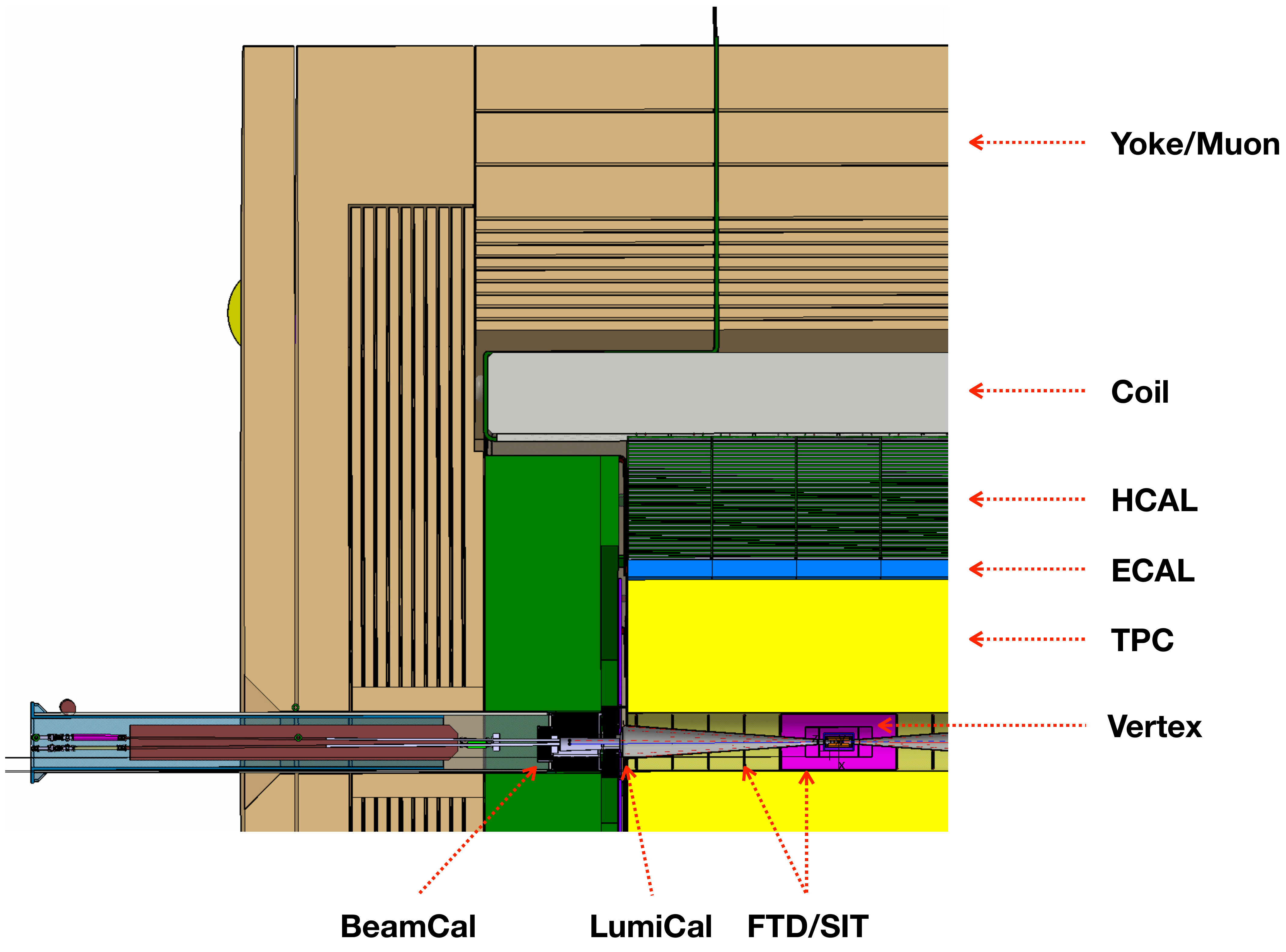} & 
 \includegraphics[width=0.35\hsize]{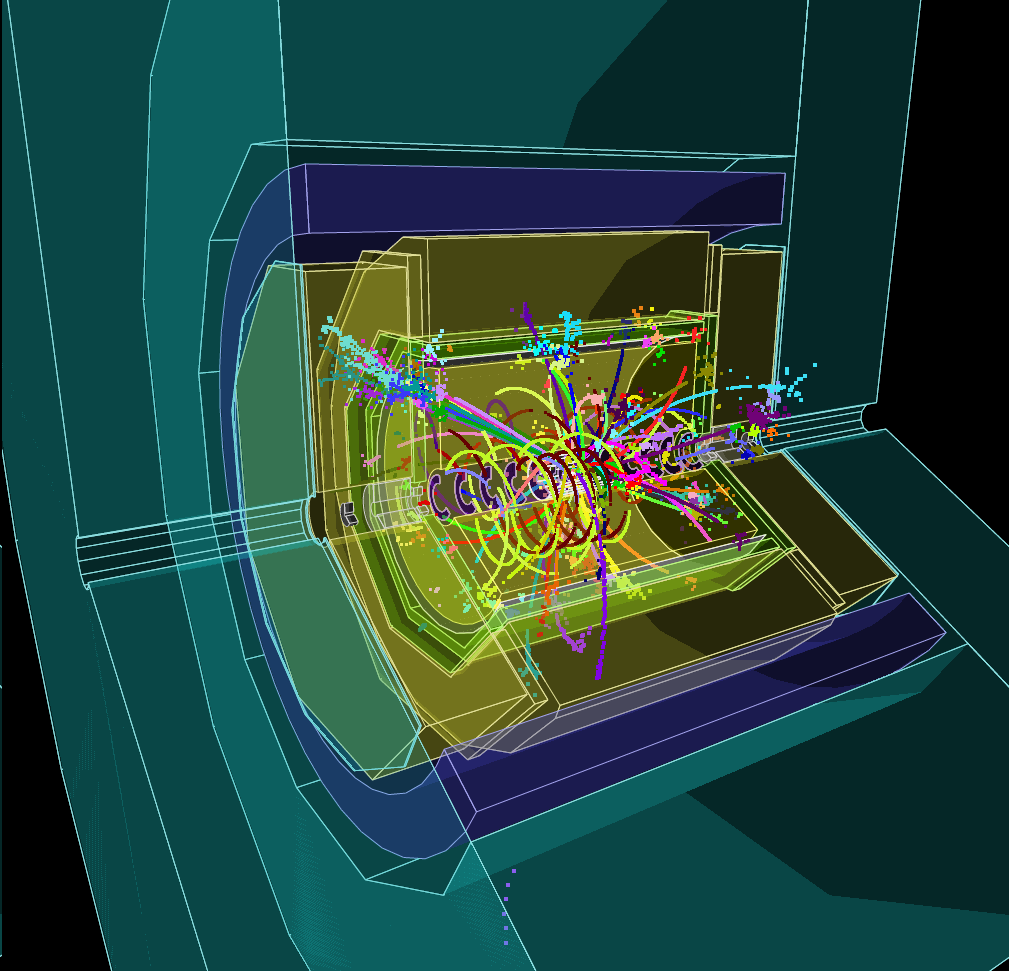}
 \\
 \end{tabular}
\caption{Left: Single quadrant view of the ILD detector. Right: Event display of a simulated hadronic decay of a $t \bar t$ event in ILD. The coloring of the tracks show the results of the reconstruction, each color corresponding to a reconstructed particle.
\label{fig:ILD}}
 \end{center}
 \end{figure}

In addition to fulfilling the science requirements, ILD has also to be able to cope with the ILC environment. Due to the extreme focusing of the two beams at the interaction region, so-called beamstrahlung is generated, which produces significant background in the detector. The magnetic field focuses the majority of the charged component of the beamstrahlung background into the forward region, but some fraction will still hit sensitive detector parts.

\section{Implementation of the ILD detector}
The ambitious requirements of the ILC detectors sparked a world-wide {{R\&D}} program to develop and demonstrate the different technologies needed~\cite{RDliaision}. The {{R\&D}} was mostly coordinated and executed within so-called {{R\&D}} collaborations, which concentrated on particular technologies and sub-detector systems. These collaborations operated outside the detector concept groups, and, in many cases, served several detector concept groups, sometimes even at more than one collider proposal. The ILD concept group from its beginning has collaborated very closely with these {{R\&D}} groups, and has organized the needed {{R\&D}} work through and with the {{R\&D}} collaborations. 

The ILD detector is a multi-purpose detector in which the different requirements are addressed by a combination of different sub-detector systems. The optimization for ultimate precision in the reconstruction of charged and neutral particles requires that all major systems are contained within a strong solenoidal magnetic field, of 3.5\,T strength. This field allows the measurement of the momentum of charged particles and removes low-energy background from the main part of the detector. Ultimate precision also requires that as little material as possible is introduced into the tracking part of the detector, which pushes in particular the coil to the outside of the tracking and calorimeter system. Over the past few years, an intense optimization effort has been undertaken by the ILD collaboration, to optimize the size of the ILD detector. Two different benchmarking models have been defined, one following the model presented in the ILD-DBD \cite{ild-dbd}, one reduced in radius by about 10\%, while keeping the length of the detector.  A three-dimensional rendering of the large detector model is shown on the title page of this document.
A quadrant view of the large detector model is shown in figure \ref{fig:ILD} (left), together with an event display in this detector of a top event (right).

The ILD concept from its inception has been open to new technologies. 
No final decision on subdetector technologies has been taken at this time, and in many cases several options are currently under consideration. In the current implementation, ILD differentiates between options and alternatives. 
For a particular technology to be accepted as an option, it has to demonstrate a certain maturity, has to demonstrate key performance parameters through test beam experiments, and has to develop a concept which addresses the main questions of integrating the technology into a subdetector for ILD. Technologies which are less mature can still be considered alternatives - meaning technologies which offer a promise for improved performance and/or lower cost - where some experimental evidence exists that this technology will eventually become a feasible option, but where the final validation is still missing. 

The main parameters of the ILD detector are summarised in table~\ref{ild:tab:barrelpara}, together with the different technological options under consideration. 

\begin{table}[th]
    \centering
    \begin{tabular}{|l|l|c|c|p{4cm}|}
    \toprule
        {\bf Technology} & {\bf Detector} & {\bf Start (mm)}   & {\bf Stop (mm)} & {\bf Comment} \\
        \midrule
pixel detectors & Vertex & $r_{in}=16$   & $r_{out}=58$   & 3 double layers of silicon pixels \\
& Forward tracking  & $z_{in}=220$ & $z_{out}=371$ & 2 Pixel disks \\
 & SIT    & $r_{in}=153$  & $r_{out}=303$  & 2 double layers of Si pixels            \\
\midrule
Silicon strip & Forward tracking  & $z_{in}=645$ & $z_{out}=2212$ & 5 layers of Si strips\\
                & SET    & $r_{in}=1773$ & $r_{out}=1776$ & 1 double layer of Si strips           \\
                & & & & \\
\midrule
Gaseous tracking & TPC & $r_{in}=329$ & $r_{out}=1770$ & MPGD readout, 220 points along the track\\
\midrule
Silicon tungsten calorimeter & ECAL option& $r_{in}=1805$ & $r_{out}=2028$ & 30 layers of $5\times 5~\mathrm{mm}^2$ pixels \\
& ECAL EC option & $z_{in}=2411$ & $z_{out}=2635$ & 30 layers of $5\times 5~\mathrm{mm}^2$ pixels \\
& Luminosity calorimeter &$r_{in}=83$ & $r_{out}=194$& 30 layers\\
& &$z_{in}=2412$ & $z_{out}=2541$& \\
\midrule
Diamond tungsten or GaAs calorimeter & Beam calorimeter &$r_{in}=18$ &$r_{out}=140$& 30 layers\\
&& $z_{in}=3115$&$z_{out}=3315$&\\
\midrule
SiPM-on-Tile & ECAL alternative   & $r_{in}=1805$ & $r_{out}=2028$ & 30 layers, 5~mm strips, crossed\\
& ECAL EC alternative& $z_{in}=2411$ & $z_{out}=2635$ & 30 layers, 5~mm strips, crossed\\
             & HCAL option   & $r_{in}=2058$ & $r_{out}=3345$ & 48 layers, $3\times 3~\mathrm{cm}^2$ pixels\\
             & HCAL EC option& $z_{in}=2650$ & $z_{out}=3937$ & 48 layers, $3\times 3~\mathrm{cm}^2$ pixels\\
\midrule
RPC          & HCAL option   & $r_{in}=2058$ & $r_{out}=3234$ & 48 layers, $1 \times 1 ~\mathrm{cm}^2$ pixels \\
& HCAL EC option & $z_{in}=2650$ & $z_{out}=3937$ & 48 layers, $1 \times 1~\mathrm{cm}^2$ pixels\\
\midrule
SiPM on scintillator bar & Muon & $r_{in}=4450$ & $r_{out}=7755$ & 14 layers \\
& Muon EC & $z_{in}=4072$ & $z_{out}=6712$ & up to 12 layers \\

%        Silicon SIT & 153& 300 & 2 layers of SI strip detectors& & & \\
%       TPC & 330 & 1808 & 220 points in TPC & & & \\
%        Silicon SET & 1811& & 2 layers of SI strip detector & & & \\
%        ECAL & 1843 & 2028 & Tungsten absorber, Si sensor, 30 layers, $5 \times 5 mm^2$ cells & 2450 & 2635 & \\
%             &      &      & Scintillator strips, crossed, 5 mm wide, 30 layers& & & \\
%        HCAL & 2058 & 3410 & RPC gas layers, $1 \times 1 cm^2 cells$, 48 layers& 2650 & 3937 & 48 layers \\
%             &      &      & SiPM on Sintillator layers, $3 \times 3 cm^2 cells$, 48 layers& & & \\
%        Muon & 4450 & 7755 & SiPM on Scintillator strips, 14 layers & 2560 & 7755 & 12 layers\\
\bottomrule
    \end{tabular}
    \caption{Key parameters of the ILD detector. All numbers from~\cite{Behnke:2013lya}. ``Star'' and ``Stop'' refer to the minimum and maxiumum extent of subdetectors in radius and/or $z$-value .}
    \label{ild:tab:barrelpara}
\end{table}

\subsection{Vertexing system}
The system closest to the interaction region is a pixel detector designed to reconstruct decay vertices of short lived particles with great precision. ILD has chosen a system consisting of three double layers of pixel detectors. The innermost layer is only half as long as the others to reduce the exposure to background hits. Each layer will provide a spatial resolution around 4~$\mu\mathrm{m}$ at a pitch of about 22~$\mu\mathrm{m}$, and a timing resolution per layer of around 2--4~$\mu\mathrm{s}$. R\&D is directed towards improving this even further, to a point which would allow hits from individual bunch crossings to be resolved.

Over the last 10 years the MAPS technology has matured close to a point where all the requirements (material budget, readout speed, granularity) needed for an ILC detector can be met. The technology has seen a first large scale use in the STAR vertex detector~\cite{ild:bib:VTXcps3}, and more recently in the upgrade of the ALICE vertex detector. 
MAPS technology in general is undergoing very rapid progress and development, with many promising avenues being explored. To minimize the material in the system, sensors are routinely thinned to 50~$\mu{\mathrm m}$. 

Other technologies under consideration for ILD are DEPFET, which is also currently being deployed in the Belle II vertex detector ~\cite{Luetticke:2017zpx}, fine pitch CCDs ~\cite{fineCCD}, and also less mature technologies such as SOI (Silicon-on-insulator) and Chronopix~\cite{RDliaision}.
Very light weight support structures have been developed, which bring the goal of 0.15\% of a radiation length per layer within reach. Such structures are now used in the Belle II vertex detector.

\begin{figure}
    \centering
    \includegraphics[width=0.45\hsize]{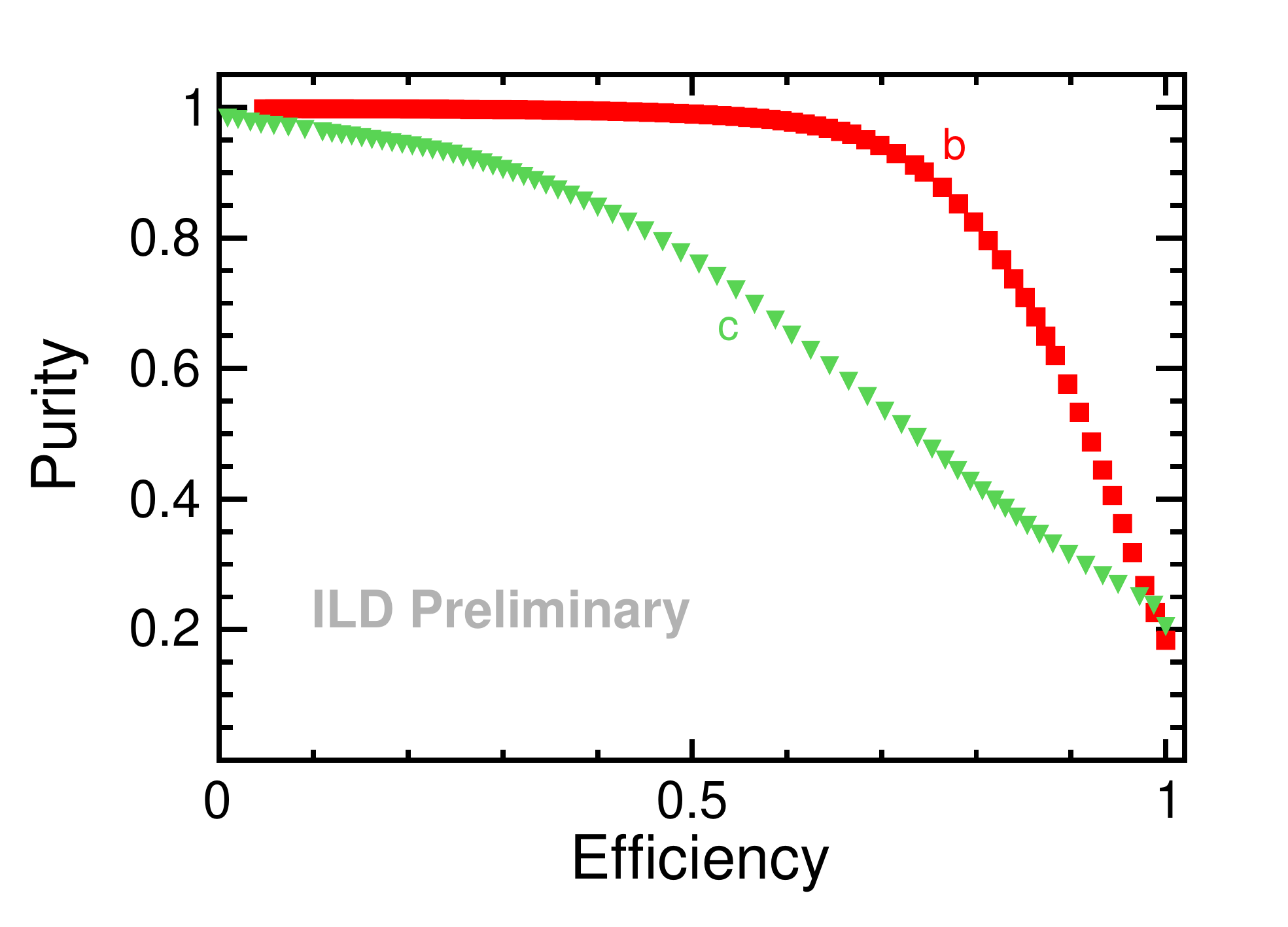}
    \includegraphics[width=0.35\hsize]{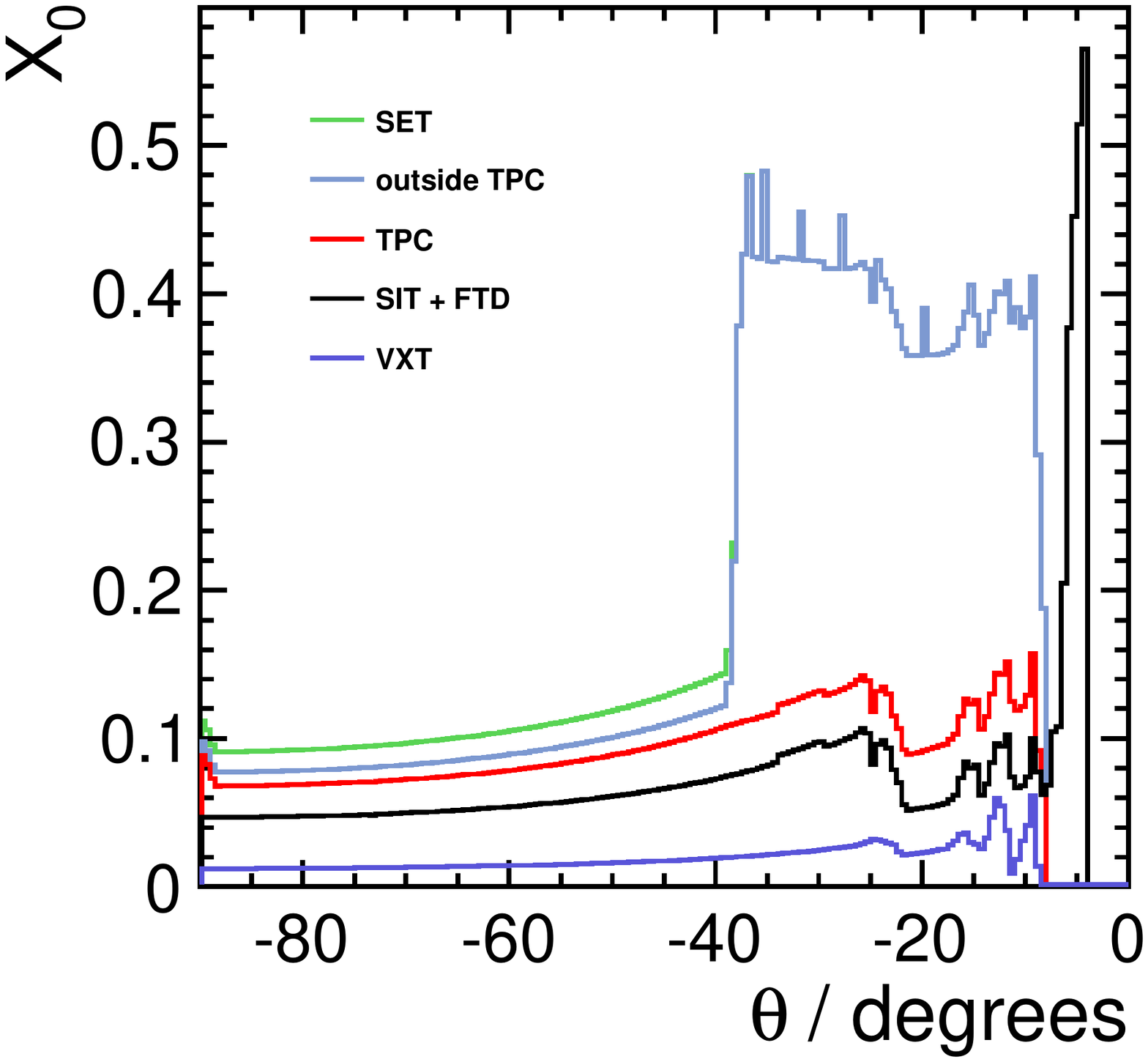}
    \caption{Left: Purity of the flavour tag as a function of the efficiency, for different flavours tagged. Right: Cumulative material budget in ILD up to the calorimeter, in fraction of a radiation length (figures from \cite{LCWS2018}).}
    \label{fig-btag}
\end{figure}

In figure~\ref{fig-btag} the purity of the flavour identification in ILD is shown as a function of its efficiency.
The performance for b-jet identification is excellent, and charm-jet identification is also good, providing a purity of about 70\% at an efficiency of 60\%.
 The system also allows the accurate determination of the charge of displaced vertices, and contributes strongly to the low-momentum tracking capabilities of the overall system, down to a few 10s of\,MeV. An important aspect of the system leading to superb flavour tagging is the small amount of material in the tracker. This is shown in figure~\ref{fig-btag} (right).

\subsection{Tracking System}

 ILD has decided to approach the problem of charged particle tracking with a hybrid solution, which combines a high resolution time-projection chamber (TPC) with a few layers of strategically placed strip or pixel detectors before and after the TPC. 
The time-projection chamber will fill a large volume about 4.6\,m in length, spanning radii from 33 to 180\,cm. In this volume the TPC provides up to 220 three dimensional points for continuous tracking with a single-hit resolution of better than 100~$\mu\mathrm{m}$ in $r \phi$, and about 1\,mm in $z$. This high number of points allows a reconstruction of the charged particle component of the event with high accuracy, including the reconstruction of secondaries, long lived particles, kinks, etc.. For momenta above 100\,MeV, and within the acceptance of the TPC, greater than 99.9\% tracking efficiency has been found in events simulated realistically with full backgrounds. At the same time the complete TPC system will introduce only about 10\% of a radiation length into the detector~\cite{Diener:2012mc}. 

\begin{figure}
    \centering
    \begin{tabular}{m{0.45\hsize}m{0.55\hsize}}
    \includegraphics[width=.9\hsize]{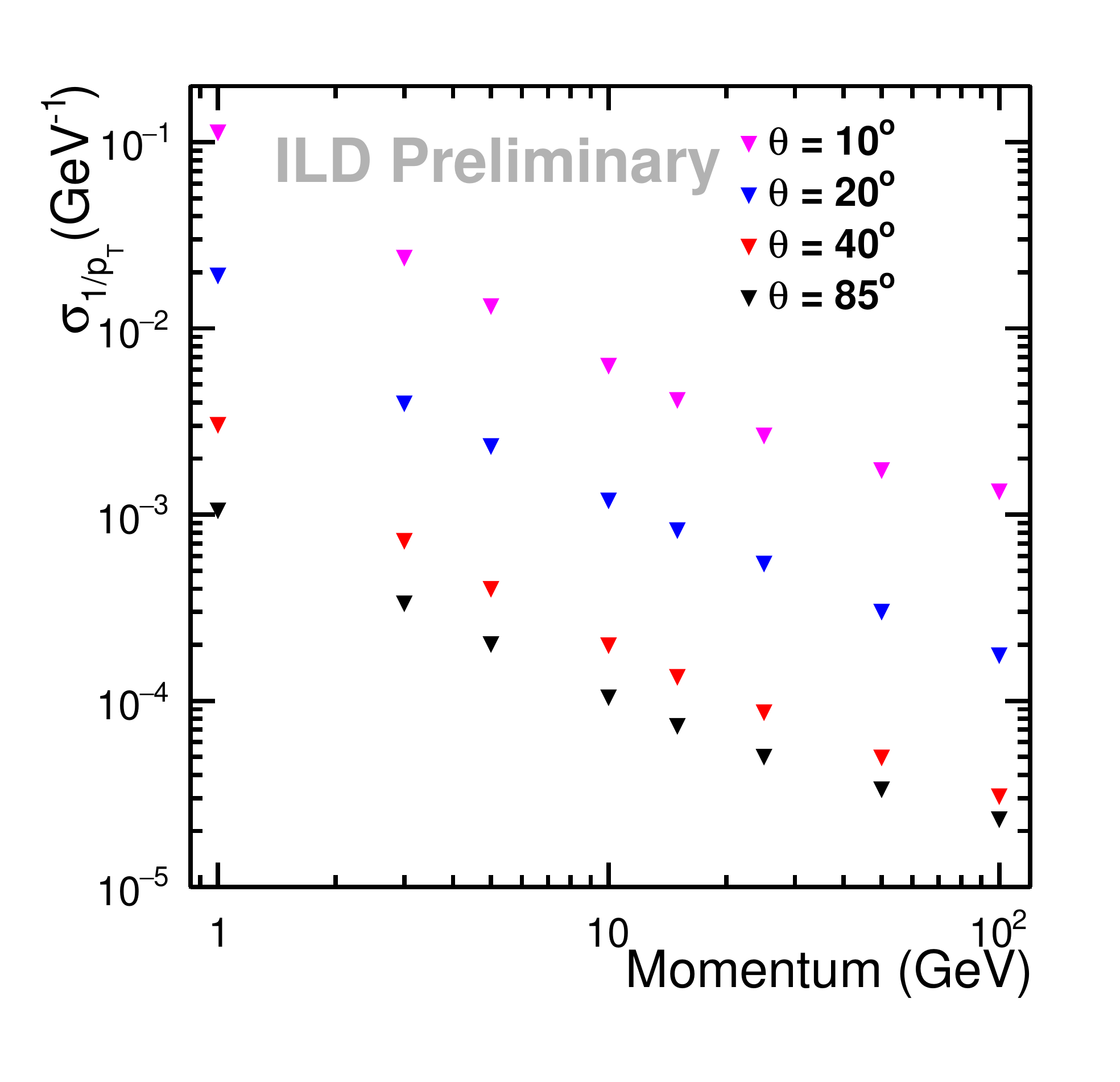} &
        \includegraphics[width=.82\hsize]{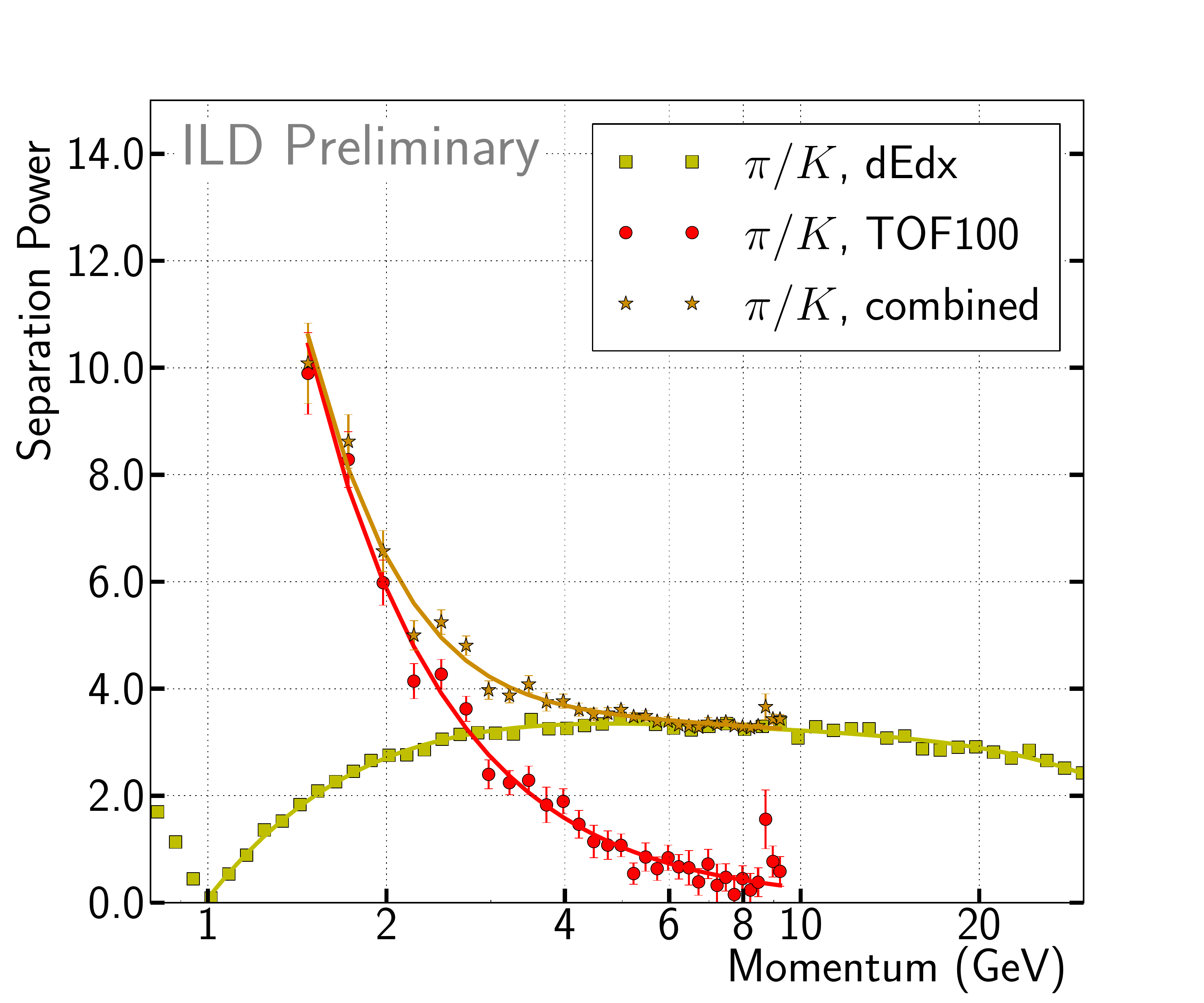}
    \end{tabular}
    \caption{ Left: Simulated resolution in $1/p_t$ as a function of the momentum for single muons. The different curves correspond to different polar angles. Right: Simulated separation power (probability for a pion to be reconstructed as a kaon) between pions and kaons, from $dE/dx$ and from timing, assuming a 100~ps timing resolution of the first ECAL layer (figures from \cite{LCWS2018}).}
    \label{fig:momentumvsp}
\end{figure}

Inside and outside of the TPC volume a few layers of silicon detectors provide additional high resolution points, at a point resolution of $10\mu \mathrm{m}$. Combined with the TPC track, this will result in an asymptotic momentum resolution of $\delta p_t / p_t^2 = 2 \times 10^{-5}$ ((GeV/c)$^{-1}$) for the complete system. Since the material in the system is very low, a significantly better resolution at low momenta can be achieved than is possible with a silicon-only tracker. The achievable resolution is illustrated in figure~\ref{fig:momentumvsp}, where the $1/p_t$-resolution is shown as a function of the momentum of the charged particle. In the forward direction, extending the coverage down to the beampipe, a system of two pixel disks (point resolution $5 \mu$m) and five strip disks (resolution $10 \mu$m outside of the TPC, and $5 \mu$m inside the TPC) provide tracking coverage down to the beam-pipe.

The time-projection chamber also enables the identification of particle type by the measurement of the specific energy loss, $dE/dx$, for tracks at intermediate momenta~\cite{Hauschild:2000eg}. The achievable performance is shown in figure~\ref{fig:momentumvsp} (right). If the inner and/or outer silicon layers can provide timing with 100~ps resolution, time of flight measurements can provide additional information, which is particularly effective in the momentum regime which is problematic for $dE/dx$, as it is shown in figure~\ref{fig:momentumvsp} (right). 

The design and performance of the TPC has been the subject of intense {{R\&D}} over the last 15 years. Several Micro Pattern Gas Detector (MPGD) technologies for the readout of the TPC have been successfully developed, and have demonstrated the required performance in test beam experiments. The readout of the charge signals is realized either through traditional pad-based systems, or by directly attaching pixel readout-ASICs to the endplate. A large volume field cage has been built to demonstrate the low mass technology needed to meet the 10\% X$_0$ goal discussed above. Most recently the performance of the specific energy loss, $dE/dx$, has been validated in test beam data. Based on these results, the TPC technology is sufficiently mature for use in the ILD detector, and can deliver the required performance (see e.g. \cite{Attie:2016yeu,Bouchez:2007pe}).

\subsection{Calorimeter System}
A very powerful calorimeter system is essential to the performance of a detector designed for particle flow reconstruction. Particle flow stresses the ability to separate the individual particles in a jet, both charged and neutral. This puts the imaging capabilities of the system at a premium, and pushes the calorimeter development in the direction of a system with very high granularity. A particularly challenging part is the precise reconstruction of the neutral hadrons in the event. A highly granular sampling calorimeter is the solution to this challenge~\cite{Sefkow:2015hna}. The conceptual and technological development of the particle flow calorimeter have been largely done by the CALICE collaboration (for a review of recent CALICE results see {e.g.} \cite{Grenier:2017ewg}). 

ILD has chosen a sampling calorimeter readout with silicon diodes as one option for the electromagnetic calorimeter. Diodes with pads of about $(5 \times 5)$ mm$^2$ are used, to sample a shower up to 30 times in the electromagnetic section. In 2018 a test beam experiment demonstrated the large scale feasibility of this technology, by showing not only that the anticipated resolution can be reached, but also by demonstrating that a sizable system can be built and operated. The test has provided confidence that scaling this to an ILD-sized system will be possible.

As an alternative to the silicon based system, sensitive layers made from thin scintillator strips are also investigated. Orienting the strips perpendicular to each other has the potential to realize an effective cell size of $5\times 5$mm$^2$, with the number of read out channels reduced by an order of magnitude. 

For the hadronic part of the calorimeter of the ILD detector, two technologies are studied, based on either silicon photo diode (SiPM) on scintillator tile technology~\cite{Simon:2010mi} or resistive plate chambers~\cite{Laktineh:2010zsa}. The SiPM-on-tile option has a  moderate granularity, with $3 \times 3$ cm$^2$ tiles, and provides an analogue readout of the signal in each tile (AHCAL). The RPC technology has a better granularity, of $1 \times 1$ cm$^2$, but provides only 2-bit amplitude information (SDHCAL). For both technologies, significant prototypes have been built and operated. Both follow the engineering design anticipated for the final detector, and demonstrate thus not only the performance, but also the scalability of the technology to a large detector. Particular challenges were the handling of the large number of channels, which requires the integration of a large part of the readout technology into the sensitive plane, and the operation of large area systems, including the handling of connected noise and timing issues. 

\begin{figure}[th]
    \centering
    \begin{tabular}{lcr}
    \includegraphics[width=0.4\hsize]{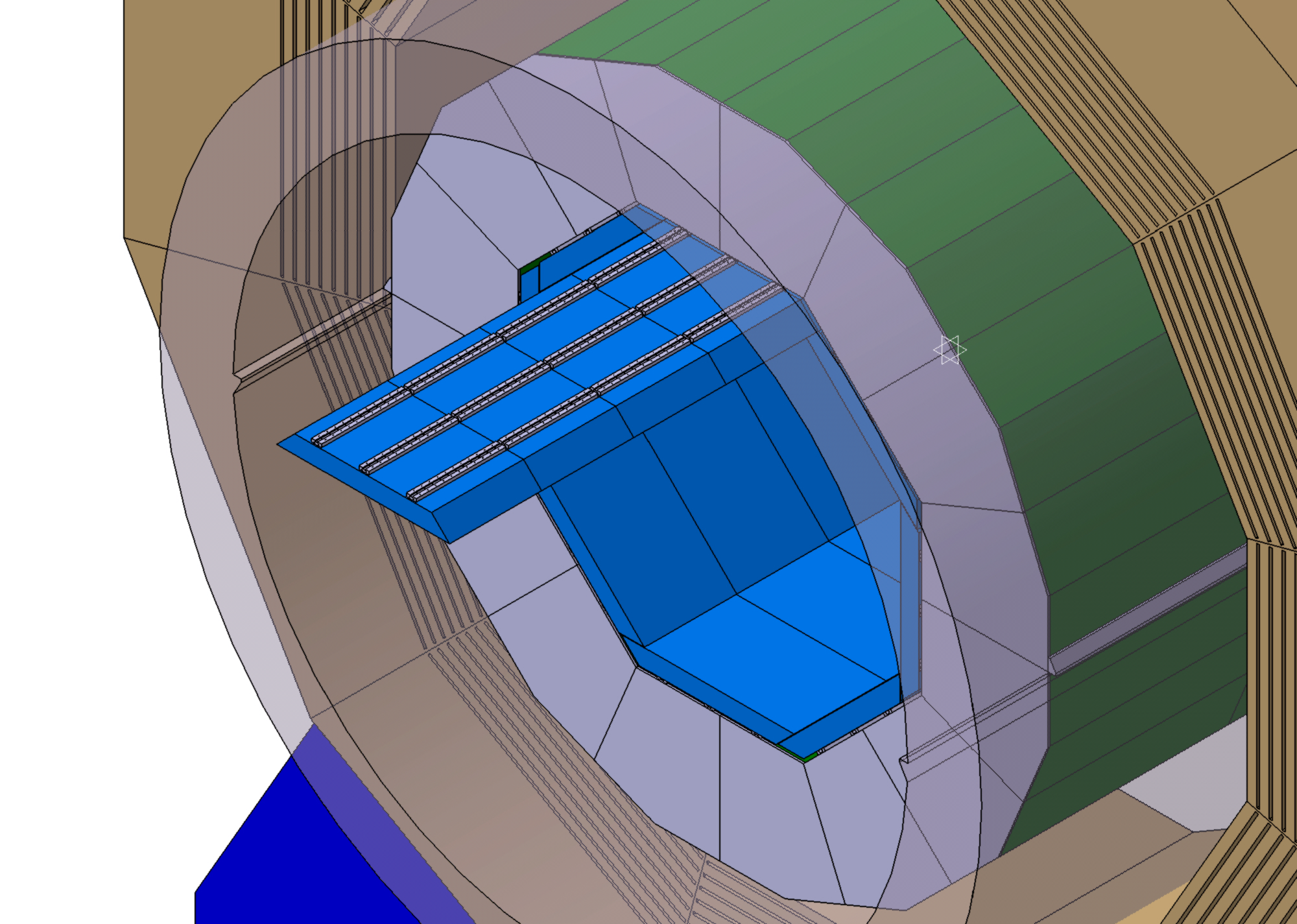} & ~~~~ &
    \includegraphics[width=0.48\hsize]{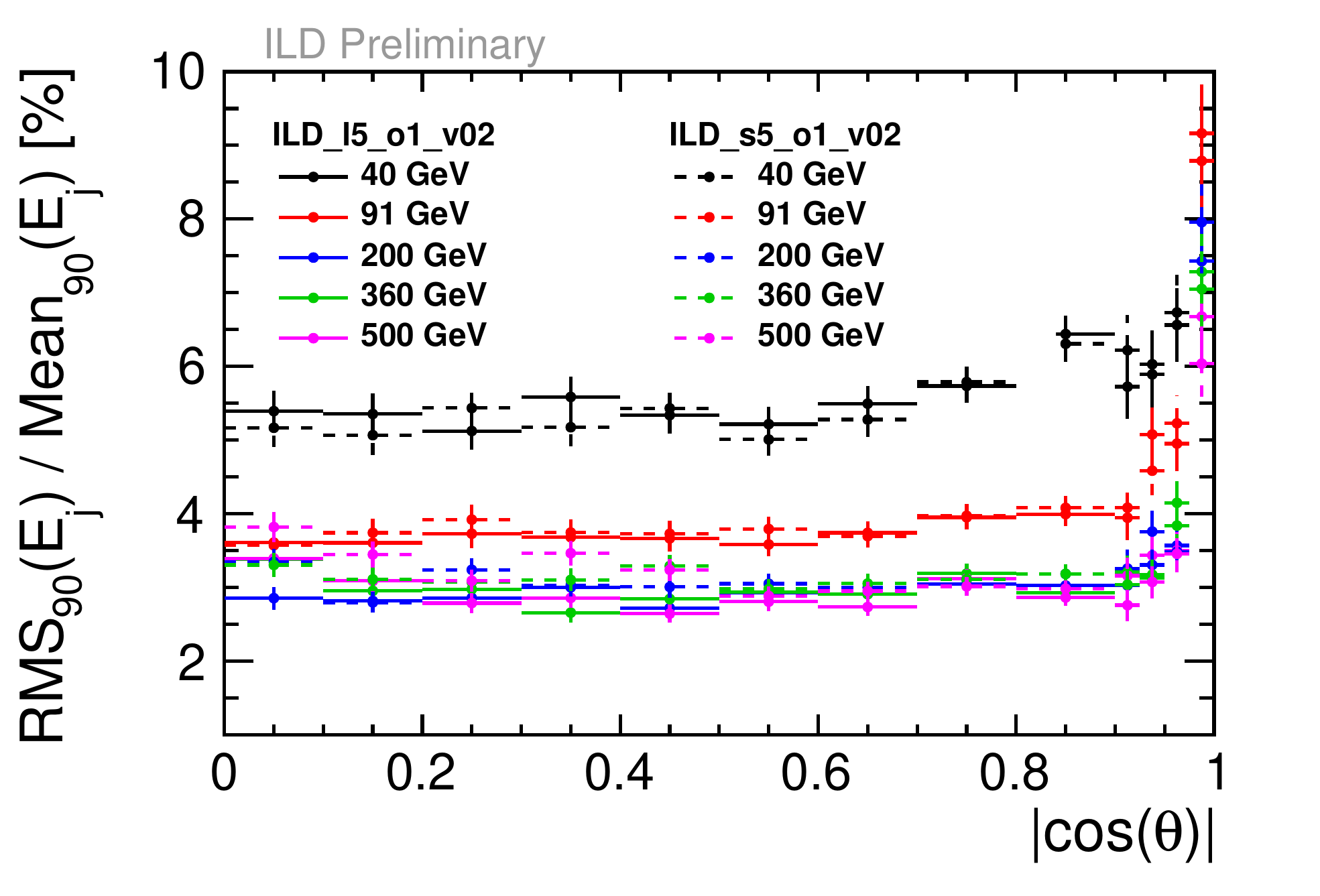}\\
    \end{tabular}
    \caption{Left: Three-dimensional rendering of the barrel calorimeter system, with one ECAL module partially extracted. Right: Particle flow performance, measured as the energy resolution in two-jet light flavour events, for different jet energies as a function of cos$(\theta)$. The resolution is defined as the rms of the distribution truncated so that 90\% of the total jet energy is contained inside the distribution. The data are shown for the large (L) and the small (S) version of the ILD detector. (figures taken from \cite{LCWS2018})}
    \label{fig:pflow}
\end{figure}

It has been a major success in the past years that the technologies needed for a true particle flow calorimeter have been successfully demonstrated in a design which is suitable for the ILD detector. With this demonstration, a major hurdle towards the realization of ILD has been overcome~\cite{Sefkow:2018rhp}. The simulated particle flow performance is shown in figure~\ref{fig:pflow}.

The iron return yoke of the detector, located outside of the coil, is instrumented to act as a tail catcher and as a muon identification system. Several technologies are possible for the instrumented layers. Both RPC chambers and scintillator strips readout with SiPMs have been investigated. Up to 14 active layers, located mostly in the inner half of the iron yoke (see table~\ref{ild:tab:barrelpara} and figure~\ref{fig:ILD} for more details) could be instrumented.

\subsection{The Forward System}
Three rather specific calorimeter systems are foreseen for the very forward region of the ILD detector~\cite{Abramowicz:2010bg}. LumiCal is a high precision fine sampling silicon tungsten calorimeter primarily designed to measure electrons from Bhabha scattering, and to precisely determine the integrated luminosity \cite{Bozovic-Jelisavcic:2014aza}. The LHCAL (Luminosity Hadronic CALorimeter) just outside the LumiCal extends the reach of the endcap calorimeter system to smaller angles relative to the beam, and closes the gap between the inner edge of the ECAL endcap and the luminosity calorimeter, LumiCal.  Below the LumiCal acceptance, where background from beamstrahlung rises sharply, BeamCal, placed further downstream from the interaction point, provides added coverage and is used to provide a fast feedback on the beam position at the interaction region. As the systems move close to the beampipe, the requirements on radiation hardness and on speed become more and more challenging. Indeed this very forward region in ILD is the only region where radiation hardness of the systems is a key requirement. In placing the different detector components, particular care has been taken to allow the bulk of the beamstrahlung photons and pairs to leave the detector through the outgoing beampipe hole. In particular the BeamCal has been positioned in such a way that backscattering of particles from the BeamCal face into the active part of the detector is minimized.

\subsection{Detector Integration and Costing}
One of the major goals of the ILD concept group was from the beginning to develop the detector concept from a collection of technological ideas to a real detector that can actually be built, commissioned, and operated within given engineering and site-dependent constraints. The effort, driven by dedicated working groups, resulted in an engineering model of ILD that describes the mechanical setup of the detector structures themselves as well as the detector services such as cabling, cooling, gas systems, and cryogenics. The technical description of ILD is based on Interface Control Documents and is documented on the ILC-EDMS system with a web-based front-end~\cite{EDMS}. A detailed CAD model of ILD exists and can be accessed at the same location.

The main mechanical structure of the ILD detector is the iron yoke that consists of three barrel rings and two endcaps. The yoke provides the required shielding for radiation and magnetic fields to allow access to the outside of the detector during data taking. The central yoke ring supports the cryostat for the detector solenoid and the barrel detectors, calorimeters and tracking system. The yoke end-caps carry the detectors in the forward direction and can be opened to allow access to the inner detector. The mechanical concept of ILD has been designed and tested in simulations for seismic conditions that can be expected at the foreseen ILC site in northern Japan.

A common concept for the detector services such as cables, cooling, gases and cryogenics has been developed. The requirements are in many cases based on engineering prototypes of the ILD sub-systems. 

The main detector solenoid is based on CMS experience and can deliver magnetic fields up to 4~T. A correction system for the compensation of the crossing angle of the ILC beam, the Detector Integrated Dipole, has been designed and can be integrated into the main magnet cryostat.

The cost of the ILD detector has been estimated at the time of the ILD detailed baseline report. The total detector cost is about $390$~Million ILCU in 2012 costs. One ILCU has been defined to be approximately equal to 1\, Dollar or 0.97 Euros in 2013. The cost of the detector is strongly dominated by the cost of the calorimeter system and the yoke, which together account for about $60\%$ of the total cost. Currently an effort is ongoing to re-evaluate the cost. In addition to the detector described in this document, and as mentioned previously, a smaller version of ILD is also considered, which will would reduce the cost by between $10$ and $20\%$. 

\section{Science with ILD}
ILD has been designed to operate with electron-positron collisions between 90 GeV and 1 TeV. The science goals of the ILC have been described in detail in \cite{ILCESU1}, and will not be repeated here. It should be pointed out that the analyses which have been performed within the ILD concept group are based on fully simulated events, using a realistic detector model and advanced reconstruction software, and in many cases includes estimates of key systematic effects. This is particularly important when estimating the reach the ILC and ILD will have for specific measurements. Determining, for example, the branching ratios of the Higgs at the percent level depends critically on the detector performance, and thus on the quality of the event simulation and reconstruction. 

In many cases the performance used in the physics analyses has been tested against prototype experiments. The key performance numbers for the vertexing, tracking and calorimeter systems are all based on results from test beam experiments. The particle flow performance, a key aspect of the ILD physics reach, could in the absence of a large scale demonstration experiment not be fully verified, but key aspects have been shown in experiments. This includes the single particle resolution for neutral and charged particles, the particle separation in jets, the linking power between tracking and calorimetry, and key aspects of detailed shower analyses important for particle flow. 

While the physics case studies are based on the version of the ILD detector presented in the detector volume of the ILC DBD~\cite{ild-dbd}, ILD has recently initiated a systematic benchmarking effort to study the performance of the ILD concept, and to determine in particular the correlations between science objectives and detector performance. The list of benchmark analyses which are under study is given in table \ref{tab-benchmark}. Even if the ILC will start operation at a center-of-mass energy of 250\,GeV, the ILD detector is being designed to meet the more challenging requirements of higher center-of-mass energies, since major parts of the detector, e.g.\ the coil, the yoke and the main calorimeters will not be replaced when upgrading the accelerator. Therefore, most of the detector benchmark analyses are performed at a center-of-mass energy of 500\,GeV, and one benchmark even at 1\,TeV. The assumed integrated luminosities and beam polarization settings follow the canonical running scenario~\cite{Barklow:2015tja}. 
In addition to the well-established performance aspects of the ILD detector, the potential of new features not yet incorporated in the existing detector prototypes, e.g.\ time-of-flight information, is being evaluated. 

The results of these studies are expected to become available in 2019 and will be published in the ILD Design Report~\cite{fwdrefIDR}. They will form the basis for the definition of a new ILD baseline detector model, which will then be used for a new physics-oriented Monte-Carlo production for 250\,GeV. Such a production is planned with the most recent beam parameters of the accelerator~\cite{Evans:2017rvt} and significantly improved reconstruction algorithms, and is expected to lead to further improvements of the expected results of the precision physics program of the ILC~\cite{ILCESU1}.

\begin{table}[thb]
    \centering
    \begin{tabular}{|p{4cm}|p{5cm}|p {5cm}|}
\hline
{\bf    Measurement}     & {\bf Main physics question} & {\bf main issue addressed} \\
\hline
Higgs mass in $H\rightarrow b {\bar b}$         &  Precision Higgs mass determination &Flavour tag, jet energy resolution, lepton momentum resolution  \\
\hline
Branching ratio $H \rightarrow \mu^+\mu^-$ & Rare decay, Higgs Yukawa coupling to muons & High-momentum $p_t$ resolution, $\mu$ identification \\
\hline
Limit on $H \rightarrow$ invisible & Hidden sector / Higgs portal & Jet energy resolution, $Z$ or recoil mass resolution, hermeticity\\
\hline
Coupling between $Z$ and left-handed $\tau$ & Contact interactions, new physics related to 3rd generation & Highly boosted topologies, $\tau$ reconstruction, $\pi^0$ reconstruction \\
\hline
$WW$ production, $W$ mass & Anomalous triple gauge couplings, $W$ mass&  Jet energy resolution, leptons in forward direction \\
\hline
Cross section of $e^+e^- \rightarrow \nu \nu qqqq$ & Vector Bosons Scattering, test validity of SM at high energies&  $W/Z$ separation, jet energy resolution, hermeticity\\
\hline
Left-Right asymmetry in $e^+e^- \rightarrow \gamma Z$ & Full six-dimensional EFT interpretation of Higgs measurements &  Jet energy scale calibration, lepton and photon reconstruction \\
\hline
Hadronic branching ratios for $H\rightarrow b \bar b $ and $c \bar c$ & New physics modifying the Higgs couplings &  Flavour tag, jet energy resolution\\

\hline
$A_{FB}, A_{LR}$ from $e^+e^- \to b\bar{b}$ and $t \bar t \rightarrow b\bar{b} qqqq / b \bar{b} qql\nu$ & Form factors, electroweak coupling &  Flavour tag, PID, (multi-)jet final states with jet and vertex charge\\
\hline

Discovery range for low $\Delta M$ Higgsinos & Testing SUSY in an area inaccessible for the LHC& Tracks with very low $p_t$, ISR photon identification, finding multiple vertices\\
\hline
Discovery range for WIMPs in mono-photon channel & Invisible particles, Dark sector & Photon detection at all angles, tagging power in the very forward calorimeters\\
\hline
Discovery range for extra Higgs bosons in $e^+e^- \rightarrow Zh$ & Additional scalars with reduced couplings to the $Z$ & Isolated muon finding, ISR photon identification.\\
\hline

%\hline
%\multicolumn{3}{|l|}{Running above the top threshold:}\\

    \end{tabular}
    \caption{Table of benchmark reactions which are used by ILD to optimize the detector performance. The analyses are mostly conducted at 500\,GeV center-of-mass energy, to optimally study the detector sensitivity. The channel, the physics motivation, and the main detector performance parameters are given.}
    \label{tab-benchmark}
\end{table}
\section{Integration of ILD into the experimental environment}
ILD is designed to be able to work in a push-pull arrangement with another detector at a common ILC interaction region. In this scheme ILD sits on a movable platform in the underground experimental hall. This platform allows for a roll-in of ILD from the parking position into the beam and vice-verse within a few hours. The detector can be fully opened and maintained in the parking position.

The current mechanical design of ILD assumes an initial assembly of the detector on the surface, similar to the construction of CMS at the LHC. A vertical shaft from the surface into the underground experimental cavern allows ILD to be lowered in five large segments, corresponding to the five yoke rings.

ILD is designed to fully cope with the ILD beam conditions. The expected levels of beam induced backgrounds have been simulated and are seen to be at tolerable levels, { e.g.} for the vertex detectors. Judiciously placed shielding keep scattered backgrounds under control. The design of the interaction region and the collimation system of the collider has been defined so as to keep the external background sources at levels below the detector requirements.

ILD is self-shielding with respect to radiation and magnetic fields to enable the operation and maintenance of equipment surrounding the detector, {e.g.} cryogenics. Of paramount importance is the possibility to operate and maintain the second ILC push-pull detector in the underground cavern during ILC operation.

\section{The ILD Concept Group}
As described above, the ILD collaboration initially started out as a fairly loosely organized group of scientists interested to explore the design of a detector for a linear collider like the ILC. With the delivery of the DBD in 2013, the group re-organised itself more along the lines of a traditional collaboration. The group gave itself a set of by-laws, which governs the function of the group, and defines rules for the membership in ILD. Groups who want to be members of ILD must sign a memorandum of participation, a first step towards an eventual memorandum of understanding to construct ILD, as soon as the ILC has been approved. 

\begin{figure}
    \centering
    \includegraphics[width=0.9\hsize]{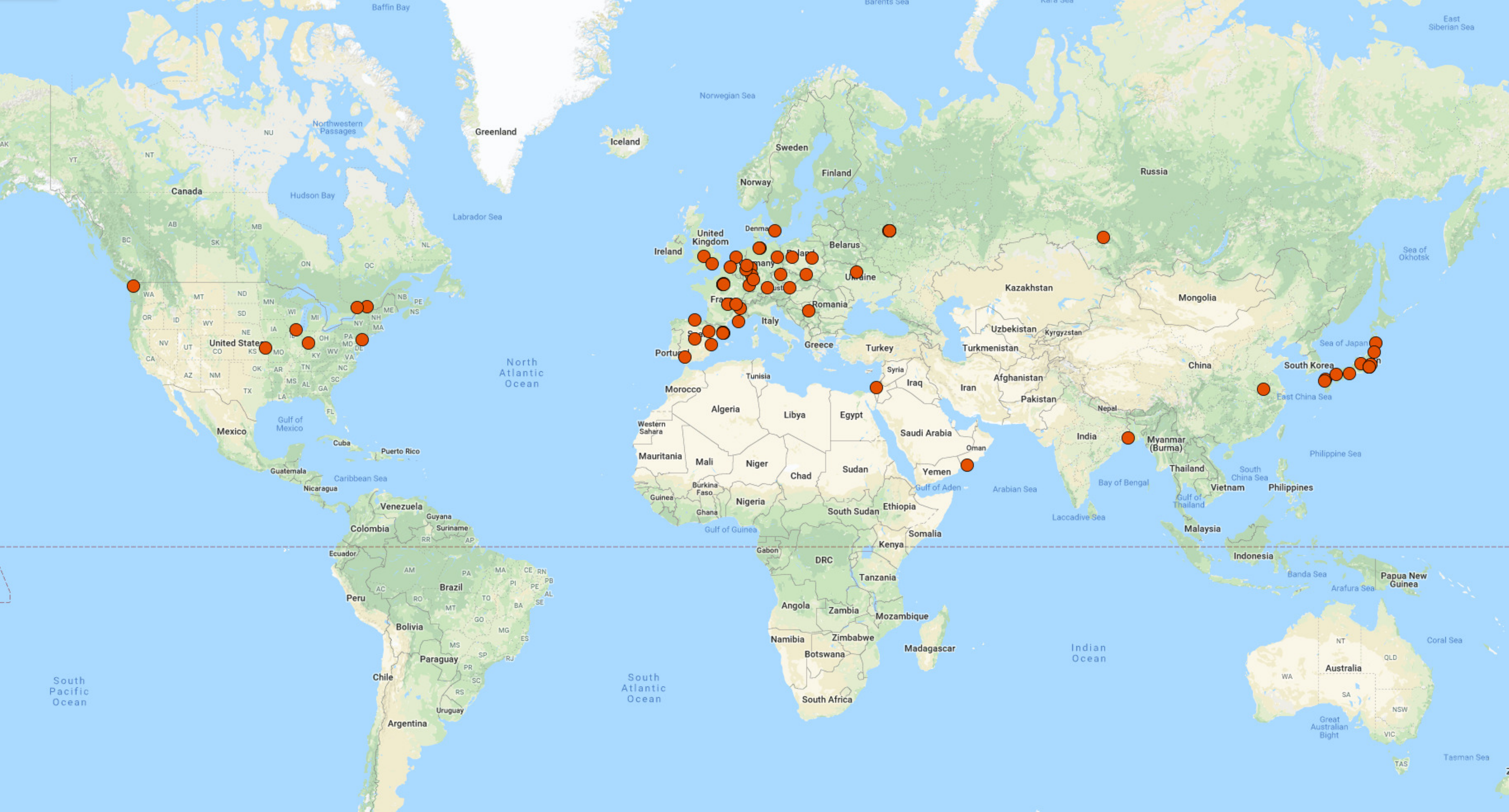}
    \caption{Map with the location of the ILD member institutes indicated.}
    \label{ild-fig-membermap}
\end{figure}

In total 61 groups from 30 countries have signed the letter of participation. A map indicating the location of the ILD member institutes is shown in figure~\ref{ild-fig-membermap}.

\section{Conclusion and Outlook}
The ILD detector concept is a well developed integrated detector optimized for use at the electron-positron collider ILC. It is based on advanced detector technology, and driven by the science requirements at the ILC. Most of its major components have been fully demonstrated through prototyping and test beam experiments. The physics performance of ILD has been validated using detailed simulation systems. A community interested in building and operating ILD has formed over the last few years. It is already sizable, encompassing 61 institutes from around the world. The community is ready to move forward once the ILC project receives approval. 

\section{References}
\bibliography{ILD}
%\begin{thebibliography}{}
%\bibitem{EDMS} ILD Technical Documentation, %http://edmsdirect.desy.de/ildtdr

%\end{thebibliography}
\end{document}